\documentclass[sii,authoryear]{ipart}

\startlocaldefs

\usepackage{longtable}
\usepackage{booktabs}
\usepackage{multirow}
\bibpunct{(}{)}{,}{autheryear}{,}{,}

\usepackage{amsmath, amsthm, amssymb, amscd, amsfonts, bm, bbm, mathrsfs}   
\usepackage{graphicx}
\usepackage{enumerate}
\usepackage[shortlabels]{enumitem}
\usepackage{url} 
\usepackage[colorlinks]{hyperref}
\usepackage[capitalize,nameinlink,noabbrev]{cleveref} 
\usepackage[dvipsnames]{xcolor}

\newtheorem{thm}{Theorem}
\newtheorem{cor}{Corollary}

\newtheorem{prop}{Proposition}

\newtheorem{eg}{Example}

\makeatletter
\newcommand*{\rom}[1]{\expandafter\@slowromancap\romannumeral #1@}
\newcommand{\HUGE}{\@setfontsize\Huge{40}{50}}   
\makeatother

\makeatletter
\newcommand{\labelcustom}[2]{%
	\protected@write \@auxout {}{\string \newlabel {#1}{{#2}{\thepage}{#2}{#1}{}} }%
	\hypertarget{#1}{#2}
}
\makeatother

\makeatletter
\newcommand{\labeltext}[3][]{%
	\@bsphack%
	\csname phantomsection\endcsname
	\def\tst{#1}%
	\def\refmarkup{}%
	\ifx\tst\empty\def\@currentlabel{\refmarkup{#2}}{\label{#3}}%
	\else\def\@currentlabel{\refmarkup{#1}}{\label{#3}}\fi%
	\@esphack%
	\labelmarkup{#2}
}
\makeatother

\makeatletter
\newcommand{\bianca}{\renewcommand\NAT@open{[}\renewcommand\NAT@close{]}}
\makeatother

\newcommand{\ed}{\overset{\mathsf{d}}{=}} 

\newcommand{\pr}{\mathsf{P}}
\newcommand{\eo}{\mathsf{E}}

\newcommand{\nd}{\mathsf{N}}

\newcommand{\ap}{\alpha} 
\newcommand{\g}{\gamma} 
\newcommand{\ga}{\Gamma} 
\newcommand{\dt}{\delta} 
\newcommand{\Dt}{\Delta} 
\newcommand{\e}{\varepsilon} 
\newcommand{\s}{\sigma} 
\newcommand{\sa}{\Sigma} 
\newcommand{\Ld}{\Lambda} 

\newcommand{\HH}{\mathbb{H}} 
\newcommand{\I}{\mathbb{I}} 
\newcommand{\N}{\mathbb{N}} 
\newcommand{\R}{\mathbb{R}} 
\newcommand{\SSS}{\mathbb{S}} 
\newcommand{\W}{\mathbb{W}} 

\newcommand{\bb}{\mathcal{B}}	
\newcommand{\sss}{\mathcal{S}}	
\newcommand{\xx}{\mathcal{X}}	
\newcommand{\yy}{\mathcal{Y}}	

\newcommand{\opnorm}[1]{
	{\vert\kern-0.25ex\vert\kern-0.25ex\vert #1 \vert\kern-0.25ex\vert\kern-0.25ex\vert}
}
\newcommand{\opnormbig}[1]{
	{\left\vert\kern-0.25ex\left\vert\kern-0.25ex\left\vert #1 \right\vert\kern-0.25ex\right\vert\kern-0.25ex\right\vert}
}

\newcommand{\opip}[1]{
	{\langle\kern-0.25ex\langle #1 \rangle\kern-0.25ex\rangle}
}
\newcommand{\opipbig}[1]{
	{\left\langle\kern-0.25ex\left\langle #1 \right\rangle\kern-0.25ex\right\rangle}
}

\let\oldnl\nl
\newcommand{\nlnonumber}{\renewcommand{\nl}{\let\nl\oldnl}}

\renewcommand{\eqref}[1]{(\ref{#1})}

\allowdisplaybreaks

\endlocaldefs

\pubyear{2022}
\volume{0}
\issue{0}
\firstpage{1}
\lastpage{1}

\begin{document}

\renewcommand{\sectionautorefname}{Section}
\renewcommand{\subsectionautorefname}{Section}
\renewcommand{\subsubsectionautorefname}{Section}

\begin{frontmatter}

\title[Inference for FoFR]{Inference for function-on-function regression: 
	central limit theorem and residual bootstrap\protect\thanksref{T1}}
\thankstext{T1}{The author is thankful to two anonymous referees for their constructive feedback.}

\begin{aug}

\author{\inits{H.}\fnms{Hyemin} \snm{Yeon}\ead[label=e1]{hyeon1@kent.edu}}
\address{Department of Mathematical Sciences, Kent State University, Kent, OH, 44242, USA\\\printead{e1}}
\end{aug}

\begin{abstract}
	We investigate asymptotic inference in a linear regression model where both response and regressors are functions,
	using an estimator based on functional principal components analysis.
	Although this approach is widely used in functional data analysis, 
	there remains significant room for developing its asymptotic properties for function-on-function regression. 
	Our study targets the mean response at a new regressor with two primary aims. 
	First, we refine the existing central limit theorem by relaxing certain technical conditions,
	which include generalizing the scaling factor,
	resulting in incorporating a broader class of random functions beyond those having scores with independence or finite higher moments.
	Second, we introduce a residual bootstrap method
	that enhances the calibration of various confidence sets for  quantities related to mean response,
	while its consistency is rigorously verified. 
	Numerical studies compare the finite sample performance of both asymptotic and bootstrap approaches, 
	demonstrating higher accuracy of the latter.
	To illustrate bootstrap inference for mean response, we apply it to the Canadian weather dataset.
\end{abstract}

\begin{keyword}[class=AMS]
\kwd[Primary ]{62R10}
\kwd{62G09}
\kwd[; secondary ]{62E20}
\end{keyword}

\begin{keyword}
\kwd{Asymptotic normality}
\kwd{Functional principal components regression estimator}
\kwd{Infinite dimensionality}
\kwd{Resampling}
\kwd{Weak convergence}
\end{keyword}


\end{frontmatter}


\newcommand{\sfA}{\mathsf{A}}
\newcommand{\BB}{\mathsf{B}}


\section{Introduction}


Over the past three decades, high-resolution measurement technologies have brought significant attention to functional data, as evidenced by numerous textbooks including \cite{RS05, FV06, SC11, HK12, HE15, KR17, CGLC24}. 
Functional regression is a key focus within functional data analysis, involving response or regressor variables as functional variables lying in an infinite-dimensional function space. 
One of the most intricate scenarios arises when both response and regressor are functions.
This case is the primary subject of this paper, termed \emph{function-on-function regression (FoFR)}.


To our knowledge, linear FoFR models were firstly studied by  \cite{RD91},
where the functional response and regressor are \emph{linearly} associated. 
This foundational work has since inspired a rich body of methodological advancements, including \cite{CFF02,MR03,YMW05b, GHK10, ISSG15, Meyer15, CYC16}.
For a review of  early developments, we refer readers to \cite{Morr15}, while more recent contributions \cite[e.g.,][]{CXC22, CFLV22} demonstrate the continued evolution of the field. 
Due to its long history, we do not attempt a comprehensive survey here.

In contrast to the rich methodological literature, 
statistical inference in linear FoFR models remains relatively underexplored,
largely due to the difficulties introduced by infinite dimensionality.  
Nonetheless, several notable efforts have been made:
one-sample hypothesis testing \citep{KMSZ08, LL25}, 
two-sample inference \citep{HKR09, XLZL22},
tests for relevant hypotheses \citep{KDD22} and for error correlation \citep{GHK10},
construction of simultaneous confidence regions \citep{Meyer15,DT24},
and central limit theorems (CLTs) \citep{CM13,BCF17}.

Notably, 
Crambes and Mas \cite{CM13} studied CLTs related to the functional principal component regression (FPCR) estimator $\hat{\BB}_{h_n}$ in FoFR.
This method utilizes the first $h_n$ functional principal components (FPCs) to approximate a functional normal equation.
Such truncation is necessary due to the finite nature of observed samples, which cannot be infinite-dimensional (cf.~\autoref{sec2}). 
This technique has been practiced for twenty years; 
refer to \cite{CFS99, CH06, CMS07, HH07, GM11, KH16a, KH16b, YDN23RB} for functional regression with a scalar response and functional regressor, known as \emph{scalar-on-function regression (SoFR)},
\cite{CM13, IK18} for instances of FoFR, 
and \autoref{sec6} for other approaches and related issues.
They found, analogously to the results as shown by \cite{CMS07} for SoFR, that 
while the FPCR slope estimator $\hat{\BB}_{h_n}$ does not weakly converge in the Hilbert-Schmidt norm, 
the estimated mean response $\hat{\BB}_{h_n}X_0$ at a new regressor $X_0$ (without considering intercept) converges in distribution to a Gaussian random function upon scaling $h_n$.
However, the CLT for the mean response in \cite{CM13} can be improved, for example, (i) by generalizing the scaling term for the mean response, as established by \cite{YDN23RB} for SoFR, and (ii) by weakening assumptions like the Bernstein inequality, which were not imposed by \cite{CMS07, YDN23RB} for SoFR.

The development of the aforementioned CLT can be a stepping stone for further research in asymptotic inference for FoFR,
such as bootstrap,
even though CLT-based statistical methods (e.g., confidence intervals calibrated by asymptotic normal distribution) may not perform well with small sample sizes.
For instance, \cite{GM11, YDN23RB} successfully developed theoretically valid bootstrap methods for the inference of the mean responses for linear SoFR,
where the latter particularly demonstrates better numerical performance of bootstrap relative to CLT. 
See \cite{YDN24PB, YDN25WB, GM11, FKV10, KH16b, RAVV16, KH24} for different bootstrap methods for various SoFR models,
which can be viewed as functional adaptations of the classical residual, paired, or wild bootstrap \citep{free81,wu86}.
We also refer to \cite{DT24, LL25} for some variants of multiplier wild bootstrap in linear FoFR models.
Yet, further asymptotic results of the FPCR estimator for FoFR remain undeveloped beyond CLT.

This paper discusses asymptotic inference of mean responses in linear FoFR models,
where the slope operator parameter $\BB$ is estimated by the FPCR estimator $\hat{\BB}_{h_n}$.
The main contributions of this paper are two-fold: (i) refining the existing CLT for mean response by \cite{CM13} and (ii) developing a residual bootstrap method for FoFR mean response while establishing its consistency. 
These advancements distinguish our work with previous studies by \cite{CMS07, CM13, YDN23RB} as follows. 
First, our approach utilizes a generalized scaling similar to \cite{YDN23RB}, now incorporating the FPCs, the new regressor $X_0$, and truncation level $h_n$.
This makes our CLT valid for a wider range of functional regressors, 
especially for the random functions with non-normal and dependent FPC scores (cf.~\autoref{eg1}).
Second, our distributional results hold conditionally on all regressors, as in  \cite{YDN23RB}.
Third, the conditions have been thoroughly examined and substantially relaxed from those in \cite{CMS07, CM13, YDN23RB}.
In particular, we only require square summability for the slope parameter, a weaker condition than absolute summability assumed by \cite{CMS07, YDN23RB}, and does not impose all finite moments as in \cite{CM13}. 
Moreover, we introduce a residual bootstrap for FoFR, adapted from \cite{YDN23RB}.
Its consistency is not only theoretically justified,
we also highlight numerical performance by contrasting coverage probabilities of the confidence sets constructed using either CLT or bootstrap, particularly for small sample sizes. 
Our results show that the bootstrap confidence sets significantly outperform their CLT counterparts, aligning with findings in \cite{YDN23RB} for SoFR.


We finalize the Introduction by outlining the structure of the paper.
After summarizing preliminary results on functional analysis and weak convergence theory for general metric spaces in \autoref{ssec_1_1}, we provide an overview of the FPCR estimator for FoFR in \autoref{sec2}.
\autoref{sec3} generalizes the CLT for the FoFR mean response upon detailing the technical conditions.
In \autoref{sec4}, we introduce a residual bootstrap procedure for FoFR and establish its theoretical validity. 
Numerical studies are given in \autoref{sec5} with focus on coverage rates,
while the Canadian weather dataset is analyzed to illustrate bootstrap inference in \autoref{secData}.
\autoref{sec6} includes concluding remarks and several future directions.
All technical details are collected in the online supplementary material.

\subsection{Mathematical preliminaries} \label{ssec_1_1}

We provide background knowledge about operator theory in general Hilbert space $\HH$ \cite[e.g.,][]{HE15, Mac09, DS88, GGK90}.
Throughout the paper, $\HH$ is assumed to be an infinite-dimensional separable Hilbert space with inner product $\langle \cdot, \cdot \rangle$ and its induced norm $\|\cdot\|$. 
A common example for $\HH$ in functional data analysis is 
\begin{align*}
	\HH 
	= L^2([0,1]) 
	\equiv \left\{ f:[0,1]\to\R|\int_0^1 f(t)^2 dt<\infty \right\},
\end{align*}
the space of all square-integrable functions supported on $[0,1]$, equipped with inner product $\langle f,g \rangle \equiv \int_0^1 f(t)g(t)dt$ for $f,g \in \HH$.
Other examples include the Sobolev spaces \cite[Section~2.8]{HE15} and the Bayes space \citep{VEP14}

Let $\bb(\HH)$ denote the class of all bounded linear operators on $\HH$. 
The sup norm $\opnorm{\cdot}_\infty$ in $\bb(\HH)$ is defined by $\opnorm{T}_\infty \equiv \sup \{ \|Bx\|: x \in \HH, \|x\|=1 \}$.
For $T \in \bb(\HH)$, if $\langle Tx, y \rangle = \langle x, Ty \rangle$ for all $x,y \in \HH$,
it is said to be self-adjoint.
A compact operator is an operator $T \in \bb(\HH)$ such that, for any bounded sequence $\{x_n\} \subseteq \HH$, $\{Tx_n\}$ has a convergent subsequence in $\HH$.
We define the tensor product $x \otimes y:\HH\to\HH$ between two elements $x,y \in \HH$ as $(x \otimes y)(z) = \langle z, x \rangle y$ for $z \in \HH$. 
If $T \in \bb(\HH)$ is self-adjoint and compact, 
it admits spectral decomposition as $T = \sum_{j=1}^\infty r_j (f_j \otimes f_j)$,
where $(r_j, f_j)$ is the $j$-th eigenelement of $T$ \citep{HE15, Mac09}.
In this case, the eigenvalues $\{r_j\}_{j=1}^\infty$ of $T$ satisfy the properties that $|r_1| \geq |r_2| \geq \cdots \geq 0$ and that $r_j \to 0$ as $j\to\infty$,
while the eigenfunctions $\{f_j\}$ of $T$ form a completely orthnormal system for the closure of image of $T$. 
Finally, $T \in \bb(\HH)$ is called a Hilbert--Schmidt (HS) operator if $\sum_{j=1}^\infty \|Tf_j\|^2 < \infty$ for a completely orthnormal system $\{f_j\}$ of $\HH$, 
and $\bb_2(\HH)$ denotes the collection of all HS operators on $\HH$, equipped with the HS norm $\opnorm{\cdot}_2$ defined as $\opnorm{T}_2 \equiv ( \sum_{j=1}^\infty \|Tf_j\|^2 )^{1/2}$.

In order to present our distributional results, we introduce the Prohorov metric \cite[cf.][]{bill99}, which is a generalization of the L\'{e}vy metric \cite[Problem~14.5]{bill95}.
To describe, let $\SSS$ denote a general metric space with metric $\rho$ and $\sss$ represent the induced Borel $\s$-algebra on $\SSS$.
For $x \in \SSS$ and $A \subseteq \SSS$, we write
$\rho(x, A) = \inf \{\rho(x,y): y \in A\}$ for the distance from $x$ to $A$,
and $A^r \equiv \{ y \in \SSS: \rho(A, y) < r \}$ is called the $r$-neighborhood of $A$ for $r \in (0,\infty)$;
if $A = \{y\}$ is a singleton, the $r$-neighborhood $A^r = B_r(x) \equiv \{ y \in \SSS: \rho(x,y)< r \}$ is the open ball of $x$ with radius $r$. 
The Prohorov metric $\pi$ on the space of all probability measures on $(\SSS, \sss)$ is then defined as
\begin{align}
	\pi(P, Q)
	& = \inf  \{ \e>0: P(A) \leq Q(A^\e) + \e,   \label{eqProhorov}
	\\& \hspace{70pt} Q(A) \leq P(A^\e) + \e, \forall A \in \sss \} . \nonumber
\end{align}
for probability measures $P,Q$ on $(\SSS, \sss)$.
Through the Prohorov metric $\pi$, we can metrize the space of probability measures, and if $\SSS$ is separable, it was shown that the weak convergence of probability measures is equivalent to the convergence with respect to $\pi$ \cite[Section~5]{bill99}.
%


\section{Estimation for linear FoFR} \label{sec2}


The FoFR model with a linear mean function is given by
\begin{align} \label{eqFOF}
	Y = \sfA + \BB X + \e,
\end{align}
where $X,Y,\e$ denote the regressor, response, and error elements.
We suppose that both random elements $X,Y$ (and hence $\e$) take values in the common Hilbert space $\HH$,
where the conditional mean of the error is zero, i.e., $\eo[\e|X] = 0$.
In this FoFR model \eqref{eqFOF}, the slope parameter $\BB$ is a bounded linear operator on $\HH$, which describes a linear relationship between two random elements $X$ and $Y$.
We call $\BB$ the \emph{slope operator} of the FoFR model \eqref{eqFOF}.
A less notable parameter $\sfA = \eo[Y] - \BB\eo[X] \in \HH$ is called the \emph{intercept element}.

The FoFR model in \eqref{eqFOF} can be deemed as a generalization of some specific cases. 
If we consider the space $L^2([0,1])$ of square integrable functions on the unit interval $[0,1]$ and the slope operator $\BB$ is the integral operator with some integrable kernel $B$ on $[0,1]^2$, i.e., $(\BB f)(t) = \int_0^1 B(s,t) f(s) ds$ for $t \in [0,1]$ and $f \in \HH$, 
then the FoFR model \eqref{eqFOF} can be written as
\begin{align} \label{eqFOFintgeral}
	Y(t) = \sfA(t) + \int_0^1 B(t,s) X(s) ds + \e(t)
\end{align}
\cite[cf.][]{CFF02}.
The model for $L^2([0,1])$ could motivate the model \eqref{eqFOF} for general Hilbert space $\HH$, 
where the latter can apply to other Hilbert space structures such as 
Sobolev spaces \cite[Section~2.8]{HE15} and the Bayes space \citep{VEP14}.

If we allow the response $Y$ (and the error $\e$) can take values in another Hilbert space, 
the FoFR model in \eqref{eqFOF} can be a generalization of the linear model with functional regressor and scalar response,
where the latter takes values in $\R$:
\begin{align} \label{eqSOF}
	Y = \ap + \langle \beta, X \rangle + \e.
\end{align}
We call it the SoFR model \citep{CFS99, CH06, CMS07, HH07, YDN23RB}.
In this case, the slope operator $\BB$ in \eqref{eqFOF} can be related to the slope function $\beta \in \HH$ in \eqref{eqSOF} as $\BB f = \langle \beta, f \rangle$ for $f \in \HH$,
while we can view the intercept $\sfA$ as a real parameter by considering the constant function with $\sfA(t) = \ap \in \R$ for all $t \in [0,1]$, when $\HH = L^2([0,1])$.

We consider an estimation for the parameters $\sfA$ and $\BB$ based on FPCs \citep{CM13, IK18}.
This estimation is motivated by normal equations.
Under the finite second moment assumptions $\eo[\|X\|^2] < \infty$ and $\eo[\|Y\|^2]<\infty$, 
we write 
\begin{align*}
	\ga & \equiv \eo[(X-\eo[X]) \otimes (X-\eo[X])], \\
	\Dt & \equiv \eo[(Y-\eo[Y]) \otimes (X-\eo[X])]
\end{align*} 
to respectively denote the covariance operator of $X$ and the cross-covariance operator between $X$ and $Y$. 
We obtain the normal equation for the FoFR model \eqref{eqFOF} as
\begin{align} \label{eq_normal_eq}
	\Dt = \BB \ga.
\end{align}
The moment equation \eqref{eq_normal_eq} can define the slope operator $\BB$, which may not be uniquely determined. 
To ensure the identifiability, we assume the following condition on $\ga$:
\begin{enumerate}[(C0)]
	\item $\ker \ga = \{0\}$. \label{condModel}
\end{enumerate}
This guarantees that the slope operator $\BB$ in the FoFR model \eqref{eqFOF} is identifiable \cite[cf.][]{CFS03, CMS07, CM13}.
Namley, if the normal equation \eqref{eq_normal_eq} hold for two operators, 
then these must be equal.


For later development, we provide expansions of $\ga$ and $X$ based on eigendecomposition \cite[e.g.,][Chapter~4]{HE15}.
Since the covariance operator $\ga$ is self-adjoint, non-negative definite, and HS (and hence compact), 
by spectral decomposition, $\ga$ admits
\begin{align*}
	\ga = \sum_{j=1}^\infty \g_j (\phi_j \otimes \phi_j),
\end{align*}
where $(\g_j, \phi_j)$ denotes the $j$-th eigenvalue-eigenfunction pair of $\ga$.
Based on this, the Karhunen--Lo\`{e}ve expansion of $X$ is written as
\begin{align*}
	X - \eo[X] = \sum_{j=1}^\infty \g_j^{1/2} \xi_j \phi_j.
\end{align*}
Here, the (normalized) FPC scores $$\{\xi_j \equiv \g_j^{-1/2} \langle X-\eo[X], \phi_j \rangle\}_{j=1}^\infty$$ are uncorrelated (but possibly dependent) with mean zero and variance one.

We observe the sample $\{(X_i, Y_i)\}_{i=1}^n$ from the FoFR model \eqref{eqFOF},
i.e.,
\begin{align} \label{eqFOFdata}
	Y_i = \sfA + \BB X_i + \e_i, \quad i=1, \dots, n,
\end{align}
for estimating the slope operator $\BB \in \bb_2(\HH)$. 
We then obtain a sample analog of normal equation \eqref{eq_normal_eq} as
\begin{align} \label{eq_normal_eq_sample}
	\hat{\Dt}_n = \BB \hat{\ga}_n + U_n.
\end{align}
In the sample normal equation \eqref{eq_normal_eq_sample},
the first two estimators 
\begin{align*}
	\hat{\ga}_n
	& \equiv n^{-1} \sum_{i=1}^n \{ (X_i-\bar{X}) \otimes (X_i-\bar{X}) \}, \\
	\hat{\Dt}_n
	& \equiv n^{-1} \sum_{i=1}^n \{ (Y_i-\bar{Y})\otimes (X_i-\bar{X}) \}, 
\end{align*}
are respectively the sample counterparts of the population covariance and cross-covariance operators, $\ga$ and $\Dt$,
while the last empirical quantity 
\[
U_n \equiv n^{-1} \sum_{i=1}^n \{(X_i - \bar{X}) \otimes (\e_i - \bar{\e})\}
\]
is the cross-covariance between the regressors and the errors, where $\bar{X} \equiv n^{-1} \sum_{i=1}^n X_i$, $\bar{Y} \equiv n^{-1} \sum_{i=1}^n Y_i$, and $\bar{\e} \equiv n^{-1} \sum_{i=1}^n \e_i$ denote the empirical averages.
Based on the sample normal equation \eqref{eq_normal_eq_sample}, it is reasonable to define an slope estimator $\hat{\BB}$ as the solution to the equation $\hat{\Dt}_n = \BB \hat{\ga}_n$.
However, unlike the population version \eqref{eq_normal_eq}, $\hat{\ga}_n$ has finite rank, meaning that its inverse cannot exist, 
when $\HH$ is infinite-dimensional.

To overcome this ill-posed inversion problem, we regularize the inversion of $\hat{\ga}_n$ by truncating the number of the FPCs used in estimation \citep{CFS99, CH06, CMS07, HH07, CM13, YDN23RB}.
Namely, we define a truncated sample inverse $\hat{\ga}_{h_n}^{-1} \equiv \sum_{j=1}^{h_n} \hat{\g}_j^{-1} (\hat{\phi}_j \otimes \hat{\phi}_j)$ to approximate the original inverse $\ga^{-1} \equiv \sum_{j=1}^\infty \g_j^{-1} (\phi_j \otimes \phi_j)$ and to subsequently obtain intercept and slope estimators $\hat{\sfA}_{h_n}$ and $\hat{\BB}_{h_n}$ of $\sfA$ and $\BB$ as
\begin{align} \label{eqFPCRest}
	\hat{\sfA}_{h_n} \equiv  \bar{Y} - \hat{\BB}_{h_n} \bar{X}, \quad \hat{\BB}_{h_n} \equiv \hat{\Dt}_n \hat{\ga}_{h_n}^{-1},
\end{align}
respectively.
The estimated mean response function $\hat{\mu}_{h_n}$ is then constructed by
\begin{align} \label{eqMRest}
	\hat{\mu}_{h_n}(x) \equiv \hat{\sfA}_{h_n} + \hat{\BB}_{h_n} x, \quad x \in \HH.
\end{align}
Here, the truncation level $h_n$ represents the number of eigenpairs from $\hat{\ga}_n$, which may grow to infinity as the sample size $n$ goes to infinity.




\section{CLT for mean response} \label{sec3}

\cite{CM13} showed that the estimator $\hat{\BB} - \BB$ cannot weakly converge in $\bb_2(\HH)$.
More precisely, if $a_n(\hat{\BB} - \BB)$ weakly converges in the Hilbert--Schmidt norm with an increasing sequence $\{a_n\}$, then the limiting distribution should be degenerate. 
Instead, they provide a CLT for mean response.
However, there is still a room for generalization/refinement.
In this section, we generalize this CLT for mean response by (i) extending the scaling, (ii) relaxing conditions, and (iii) providing conditional results given all regressors.

To discuss asymptotic distributional results for mean response,
let $(X_0, Y_0)$ denote a new regressor-response pair that is an independent copy of $(X, Y)$ and independent of the data sample $\{(X_i, Y_i)\}_{i=1}^n$. 
We are interested in the inference of the (conditional) mean response $\mu(X_0) \equiv \eo[Y_0|X_0] = \sfA + \BB X_0$ at a new predictor $X_0$. 
Our goal is to study asymptotic distributional properties of the estimated mean response $\hat{\mu}_{h_n}(X_0) \equiv \hat{\sfA}_{h_n} + \hat{\BB}_{h_n} X_0$ from \eqref{eqMRest} at $X_0$. 

Before expanding our theoretical development, 
we provide generalized scaling terms for a CLT in FoFR model \cite[cf.][]{CMS07, YDN23RB}. 
A CLT for the estimated conditional mean $\hat{\mu}_{h_n}(X_0)$ involves a scaling term 
\begin{align} 
	t_{h_n}(x) 
	& \equiv \sum_{j=1}^{h_n} \g_j^{-1} \langle x - \eo[X], \phi_j \rangle^2, \quad x \in \HH, \label{eq_scaling_t}
	\\& = \|\ga_{h_n}^{-1/2} (x-\eo[X]) \|^2,  \nonumber
\end{align}
where $\ga_{h_n}^{-1/2} \equiv \sum_{j=1}^{h_n} \g_j^{-1/2} (\phi_j \otimes \phi_j)$ is a truncated version of the square-root inverse covariance operator $\ga^{-1/2} \equiv \sum_{j=1}^\infty \g_j^{-1/2} (\phi_j \otimes \phi_j)$. 
The scaling \eqref{eq_scaling_t} can approximate 
\begin{align*}
	\|\ga^{-1/2} (x-\eo[X])\|^2 = \sum_{j=1}^\infty \g_j^{-1} \langle x-\eo[X], \phi_j \rangle^2,
\end{align*}
which is the squared reproducing kernel Hilbert space norm involved with the covariance operator $\ga$ \citep{wahba73}.
This is related to the tuning parameter $h_n$ used in the regression estimator $\hat{\BB}_{h_n}$ in \eqref{eqFPCRest}.
Its sample counterpart is defined as
\begin{align} 
	\hat{t}_{h_n}(x) 
	& \equiv \sum_{j=1}^{h_n} \hat{\g}_j^{-1} \langle x-\bar{X},  \hat{\phi}_j \rangle^2 \quad x \in \HH, \label{eq_scaling_t_est}
	\\& = \|\hat{\ga}_{h_n}^{-1/2} (x-\bar{X}) \|^2,  \nonumber
\end{align}
by using the finite-sample approximation $\hat{\ga}_{h_n}^{-1/2} \equiv \sum_{j=1}^{h_n} \hat{\g}_j^{-1} (\hat{\phi}_j \otimes \hat{\phi}_j)$ of $\ga_{h_n}^{-1}$ constructed based on the data regressors $\{X_i\}_{i=1}^n$.
Refer to \cite{CMS07,YDN23RB} for further discussion on these scaling terms in \eqref{eq_scaling_t}--\eqref{eq_scaling_t_est}

\autoref{ssec_3_1} devotes to describing technical assumptions, which are weakened compared to those imposed by the previous studies,
while we provide our refined CLT in \autoref{ssec_3_2}.

\subsection{Assumptions} \label{ssec_3_1}


We consider the following conditions, which are similar to but weaker than those in the previous work \citep{CMS07, CM13, YDN23RB}.
\begin{enumerate}[(\text{C}1)]
	\item $\BB \in \bb_2(\HH)$ such that the sequence $$\left\{ \sum_{j'=1}^\infty \langle \BB \phi_j, \phi_{j'} \rangle^2 \right\}_{j=1}^\infty$$ is non-increasing; \label{condSlope}
	
	\item $\sup_{j \in \N} \g_j^{-2} \eo[ \langle X - \eo[X], \phi_j \rangle^4]<\infty$; \label{condMoment}

\end{enumerate}

The first two conditions serve as both clarifying and relaxing the assumptions in the existing relevant papers. 
Condition~\ref{condSlope} on the slope operator is quite mild as in \cite{CM13}; but the authors overlooked the monotonicity condition in \ref{condSlope}, although it can give a critical counter example when using the Abel's theorem (cf.~Lemma~S12 
in the online supplementary material) as suggested \citet[Section~173]{Hardy21}.
Under the SoFR model \eqref{eqSOF} with $\BB = \langle \beta, \cdot \rangle$, this condition can be reduced to $\beta \in \HH$, 
implying that our results relax the absolute summability condition $\sum_{j=1}^\infty |\langle \beta, \phi_j \rangle|<\infty$ imposed in the previous works by \cite{CMS07, YDN23RB}.

Condition~\ref{condMoment} ensures the finite fourth moment of the regressor $X$, while similar or stronger moment conditions are commonly imposed in the literature \citep{CH06, CMS07, HH07, CM13}.
In particular, with Condition~\ref{condMoment}, we can generalize the weak convergence result in \cite{CM13} to a wider class of regressor functions.
To explain, \cite{CM13} assume a Bernstein's inequality to the independent FPC scores (see Section~1.4 therein).
This is fairly stronger than Condition~\ref{condMoment},
because it requires all finite moments along with independence between all the FPC scores.
\autoref{eg1} illustrates cases satisfying Condition~\ref{condMoment} but violating all finite moments as well as independence.

\begin{enumerate}[(C3)]
	\item the eigenvaules $\{\g_j\}_{j=1}^\infty$ satisfy $\g_j = \varphi(j)$ for sufficiently large $j$, where $\varphi:(0,\infty) \to (0,\infty)$ is a convex function;\label{condEVconvex}
\end{enumerate}

\begin{enumerate}[(C4)]
	\item $\sup_{j \in \N} \g_j j \log j <\infty$; and \label{condEVdecay}
\end{enumerate}

\begin{enumerate}[(C5)]
	\item $h_n^{-1} + n^{-1/2} h_n^2 \log h_n + n^{-1} \sum_{j=1}^{h_n} \dt_j^{-2} \to 0$ as $n\to\infty$, where $\{\dt_j\}$ are eigengaps defined as $\dt_j \equiv \g_1 - \g_2$ if $j=1$ and $\dt_j \equiv \min \{\g_j - \g_{j+1}, \g_{j-1} - \g_j \}$ if $j > 1$ otherwise. \label{condBasicTrunc}
\end{enumerate}

The rest of the conditions \ref{condEVconvex}-\ref{condBasicTrunc} are related to the decay rates of the eigenvalues $\{\g_j\}$ and eigengaps $\{\dt_j\}$.
The convexity condition in \ref{condEVconvex} covers standard decay rates such as polynomial or exponential, which implies $\dt_j \equiv \g_j - \g_{j+1}$.
Condition~\ref{condEVconvex} is the key condition to derive important convexity lemmas used for proving fundamental perturbation theorems,  pioneered by \cite{CMS07}.
Conditions~\ref{condEVdecay}-\ref{condBasicTrunc} are mild decay rate conditions that are relevant to technical details in perturbation theory.
These conditions are ignored by the previous works \cite[e.g.,][]{CMS07, CM13},
but it turns out that these are needed to handle subtle technical issues in CLTs \cite[cf.][]{YDN23RB, YDN24PB}.
For more discussion on the last term $n^{-1} \sum_{j=1}^{h_n} \dt_j^{-2}$ in Condition~\ref{condBasicTrunc}, refer to Lemma~S2 in the supplement of \cite{YDN23RB} and the paragraph that comes before it or Remark~S1 
of the online supplementary material.

To treat  bias from truncation in the limiting distribution, 
we suppose the following balancing condition between the decay rates of $\{\dt_j\}$ and $\{\|B\phi_j\|\}$:
\begin{enumerate}[$B(u)$]
	\item $\sup_{j \in \N} j^{-1} m(j,u) \|B\phi_j\|^2 < \infty$, \label{condBias}
\end{enumerate}
for constant $u \in (0,\infty)$ and function
\begin{align}
	m(j,u) \equiv \max \left\{ j^u, \sum_{l=1}^j \dt_l^{-2} \right\}. \label{eq_mju}
\end{align}
This is adapted from the previous works \citep{CMS07, YDN23RB} on the SoFR fit to the FoFR, 
and will be used to handle the bias term related to $\BB\Pi_{h_n} - \BB$. 
Here, for each positive integer $h$, $\Pi_h \equiv \sum_{j=1}^h (\phi_j \otimes \phi_h)$ denotes the projection operator onto the linear space spanned by first $h$ eigenfunctions $\{\phi_j\}_{j=1}^h$.

It is common to consider more explicit structures on the decay rates.
In the following, for generic sequence of positive real numbers $\{r_n\}$ and $\{s_n\}$, we write $r_n \asymp s_n$ if $r_n/s_n$ is bounded away from both zero and infinity. 
The polynomial decay rates in our framework are then stated as follows:
\begin{enumerate}[(P)]
	\item $\dt_j \asymp j^{-a}$ and $\|\BB\phi_j\| \asymp j^{-b}$ for $a>2$ and $b>5/2$ with $a/2+1 < b$. \label{condPoly}
	
\end{enumerate}
Such decay rates commonly appear in functional data literature such as \cite{CH06, HH07, KH16a, lei14, YDN23RB, LL25},
and the rates in Condition~\ref{condPoly} are quite mild compared to those in the previously mentioned papers.
This condition will simplify conditions for our CLT below
since it may imply Conditions~\ref{condEVconvex}-\ref{condBasicTrunc} and \ref{condBias}.

For the weak convergence for the estimated mean response $\hat{\mu}_{h_n}(X_0) \equiv \hat{\sfA}_{h_n} + \hat{\BB}_{h_n}X_0$, we also consider the following mild assumption on the scaling term $t_{h_n}(X_0)$ from \eqref{eq_scaling_t}:
\begin{enumerate}[(S)]
	\item $h_n t_{h_n}(X_0)^{-1} = O_\pr(1)$, \label{condScale}
\end{enumerate}
which means the scaling $t_{h_n}(X_0)$ can scale more than $h_n$.
Important examples of Condition~\ref{condScale} include the FPC scores with $h_n t_{h_n}(X_0)^{-1}$ converges to some random variable $V$ with $\pr(0<V<\infty) = 1$ in probability as $n\to\infty$.
In particular, if the FPC scores are independent, which is assumed in \cite{CM13}, $t_{h_n}(X_0)$ and $h_n$ are asymptotically equivalent in the sense that $h_n t_{h_n}(X_0)^{-1} \xrightarrow{\pr} 1$.
Thus, in this special case, our weak convergence result can be reduced to Theorem~9 of \cite{CM13}.

In this work, we assume that the error element $\e$ is essentially homoscedastic with constant covariance operator $\eo[\e\otimes\e|X] \equiv \sa \in \bb(\HH)$. 
More precisely, we impose the following condition that is adapted from \cite{YDN23RB}.
\begin{enumerate}[(E)]
	\item There exists a class $\{g_x\}_{x \in \HH}$ of (non-random) functions on $[0,\infty)$ to $[0,\infty)$ with $\lim_{u\to\infty}g_x(u) = 0$ for each $x \in \HH$ such that  $\eo[\langle \e, x \rangle^2 \I(|\langle \e, x \rangle| \geq u)|X] \leq g_x(u)$ almost surely for any $u \geq 0$ and $x \in \HH$. \label{condError}
\end{enumerate}
Condition~\ref{condError} is mild and includes a trivial but important case when the error $\e$ and the regressor $X$ are independent.

Finally, for better exposition, we provide an example that can satisfy all the assumptions described above.
Its construction itself is relatively simple, although its nature is quite complicated, which can be shown through the non-normal and dependent FPC scores of the functional regressor. 
See \cite{YDN23RB, YDN24PB} for similar examples.

\begin{eg} \label{eg1}
	
	Suppose that $\HH = L^2([0,1])$ with the eigenfunctions $\{\phi_j\}$ being the Fourier basis functions,
	implying Condition~\ref{condModel}.
	Suppose that the underlying regressor $X$ is generated as follows:
	\begin{enumerate}[(D)]
		\item  the regressor function $X$ has the (normalized) FPC scores as $\g_j^{-1/2} \langle X, \phi_j \rangle = W_j\xi$ for $j=1,2,\dots,$, 
		where $\{W_j\}$ is an iid sequence independent of the latent variable $\xi$ and the variables $W_j, \xi$ all have zero mean and unit variance with $\pr(\xi=0)=0$.  \label{egFPCdep}
	\end{enumerate}
	The regressor from Condition~\ref{egFPCdep} satisfies Condition~\ref{condScale}, as $h_n^{-1} \sum_{j=1}^{h_n} \xi_j^2 \xrightarrow{\pr} \xi^2$ holds due to the construction in \ref{egFPCdep}.
	We could further assume that finite fourth moments as $\sup_{j \in \N} \eo[W_j^4]<\infty$ and $\eo[\xi^4]<\infty$, which guarantees Condition~\ref{condMoment}.
	For instance, $W_j$ are iid normal variable while the common variable $\xi$ follows the centered $\mathsf{t}(\nu)$-distribution with $\nu>4$. 
	Upon imposing the polynomial decay rates in \ref{condPoly}, 
	the conditions related to the decay rates are satisfied. 
	By considering the error independent of the regressor, this construction satisfies all the conditions given in this section.
	
\end{eg}

\subsection{Generalized CLT} \label{ssec_3_2}


We provide our generalized CLT for mean response in the FoFR model \eqref{eqFOF}.
Our goal is to study the asymptotic distribution of the following quantity:
\begin{align}
	T_n = T_n(X_0) 
	& \equiv \sqrt{n \over \hat{t}_{h_n}(X_0)} \{\hat{\mu}_{h_n}(X_0) - \mu(X_0)\}, \label{eqStat}
\end{align}
where $\hat{\mu}_{h_n}(X_0) \equiv \hat{\sfA}_{h_n} + \hat{\BB}_{h_n} X_0$ from \eqref{eqMRest} and $\mu(X_0) \equiv \sfA + \BB X_0$ are the mean responses at the sample and population levels, respectively.
In what follows, 
$\mathsf{G}(\rho, \Ld)$ represents the Gaussian distribution on $\HH$ with mean element $\rho \in \HH$ and covariance operator $\Ld:\HH\to\HH$.
Namely, for a Gaussian element $Z \sim \mathsf{G}(\rho, \Ld)$, its projection $\langle Z, x \rangle$ onto each $x \in \HH$ is a normal random variable with mean $\langle \rho, x \rangle$ and variance $\langle \Ld x, x \rangle$.
Also, let $\xx_n \equiv \{X_i\}_{i=1}^n$ denote the set of all observed regressor functions. 
The following theorem shows the refined CLT for mean response, 
where its proof is given in Section~S3 
of the online supplementary material.

\begin{thm} \label{thmCLT}
	Suppose Conditions~\ref{condModel}-\ref{condBasicTrunc} and \ref{condError} hold along with Condition~\ref{condScale} and 
	\begin{align} \label{eqConvRate_h_upper}
		h_n^{-1} + n^{-1/2} h_n^{5/2} (\log h_n)^2 \to 0.
	\end{align}
	We additionally suppose that Condition~\ref{condBias} holds for $u>5$ along with 
	\begin{align} \label{eqConvRate_h_lower}
		n = O(m(h_n, u)),
	\end{align}
	where the function $m$ is given in \eqref{eq_mju}.
	Then, the mean response statistic $T_n$ in \eqref{eqStat} weakly converges to the centered Gaussian distribution $\mathsf{G}(0,\sa)$ with covariance $\sa$ conditionally on $\xx_n \equiv \{X_i\}_{i=1}^n$ and $X_0$.
	More precisely, denoting $P_n \equiv \pr(T_n \in \cdot|\xx_n, X_0)$ for the conditional distribution of $T_n$ given $\xx_n$ and $X_0$ and $P$ for $\mathsf{G}(0,\sa)$, 
	we have 
	\begin{align*}
		\pi(P_n, P) \xrightarrow{\pr} 0.
	\end{align*}
		%
		%
\end{thm}

\autoref{thmCLT} generalizes and substantially reinforces the CLT results in the previous works by \cite{CMS07, YDN23RB, CM13} in the following perspectives. 
First, the generalized scaling $t_{h_n}$ in \eqref{eq_scaling_t} (or $\hat{t}_{h_n}$ in \eqref{eq_scaling_t_est}) can include a variety of classes of regressor functions.
The functional regressors with non-normal and dependent FPC scores are an important example, where the results in \cite{CM13} may not be able to apply this case (cf.~\autoref{prop_counter}).
Second, the asymptotic normality results in \autoref{thmCLT} hold conditionally on all the regressors $\xx_n \equiv \{X_i\}_{i=1}^n$ and $X_0$, similarly to the CLT for SoFR by \cite{YDN23RB}, 
while \cite{CM13} provided only unconditional CLTs. 
Lastly, we relax many assumptions in these previous works: all finite moments and independence by \cite{CM13}, the asymptotic equivalence between $h_n$ and $t_{h_n}(X_0)$ (i.e., $h_n^{-1} t_{h_n}(X_0) \xrightarrow{\pr} 1$) by \cite{CMS07}, and the absolute summability of the slope by \cite{CMS07, YDN23RB}.

We provide a subsequent CLT result under polynomial decay rates in Condition~\ref{condPoly}.
This can replace the truncation conditions given in 
\eqref{eqConvRate_h_upper} and \eqref{eqConvRate_h_lower} by a simpler form, as described next.

\begin{cor} \label{corCLT}
	Suppose Conditions~\ref{condModel}, \ref{condMoment}, \ref{condScale}, \ref{condError}, and \ref{condPoly} hold along with $n \asymp h_n^u$ with $u \in (5, a+2b-1)$, where the decay rates $a>2$ and $b>5/2$ are given in Condition~\ref{condPoly}.
	Then, the CLT for $T_n$ from \eqref{eqStat} in \autoref{thmCLT} still holds.
	
	
\end{cor}

The most crucial distinction of our CLTs (\autoref{thmCLT}) from the existing CLTs in \cite{CM13} is the scaling.
Specifically, the CLTs with our scaling in either \eqref{eq_scaling_t} or \eqref{eq_scaling_t_est} can be applied to a wider range of function classes.
In particular, the existing CLTs by \cite{CM13} may not be valid for functional data with (non-Gaussian) dependent FPC scores, 
which are often found in real-world data examples \cite[cf.][]{YDN23RB}.
To further clarify this refinement, we next present an example that highlights the considerable impact of the dependence among the FPC scores on the CLTs in the FoFR model \eqref{eqFOF}.

\begin{prop} \label{prop_counter}
	Consider the example in \ref{egFPCdep} with $\pr(W_1 = 1) = 1/2 = \pr(W_1 = -1)$ and $\xi \sim \nd(0,1)$; more generally, the conditions $\sup_{j \in \N} \eo[W_j^4]<\infty$ and $\eo[\xi^4]<\infty$ are sufficient.
	Then, under the assumptions in \autoref{thmCLT}, as $n\to\infty$,  the mean response statistic upon scaling $h_n$
	converges in distribution to $|\xi_0|Z_0$,
	i.e.,
	\begin{align*}
		\sqrt{n \over h_n}\{\hat{\mu}_{h_n}(X_0) - \mu(X_0)\} \xrightarrow{\mathsf{d}} |\xi_0 |Z_0,
	\end{align*}
	where the variable $\xi_0\ed \xi$ and element $Z_0 \sim \mathsf{G}(0, \sa)$ are independent.
\end{prop}



\section{Residual bootstrap} \label{sec4}

We turn our attention to bootstrap inference of conditional mean in the FoFR model \eqref{eqFOF}. We propose a residual bootstrap method, akin to the one used in the SoFR as presented by \cite{YDN23RB},
to approximate the distribution of the mean response statistic $T_n$ in \eqref{eqStat}.
Contrary to bootstrap methods for finite-dimensional regression problems \cite[e.g.,][]{Efron79,free81}, a tuning parameter is involved in estimation for functional linear models. 
This allows us to use different tuning parameters, which can enhance the performance of bootstrap inference with more flexibility.
In particular, for our residual bootstrap procedure, we incorporate three tuning parameters as described next.

We generate bootstrap errors $\e_1^*, \dots, \e_n^*$ by drawing uniformly from the (centered) residuals $\{ \hat{\e}_{i,k_n} \}_{i=1}^n$ with replacement.
These residuals are computed as $\hat{\e}_{i,k_n} \equiv Y_i - \hat{\mu}_{k_n}(X_i)$, 
where the estimated mean response operator $\hat{\mu}_{k_n}$ is from \eqref{eqMRest} but with the initial truncation $k_n$.
We use this least consequential truncation $k_n$ only for computing the residuals. 
Next, we calculate the bootstrap responses as 
\begin{align} \label{eqFOFbts}
	Y_i^* = \hat{\sfA}_{g_n} + \hat{\BB}_{g_n} X_i + \e_i^*, \quad i=1, \dots, n
\end{align}
akin to the data generation in \eqref{eqFOFdata}.
Here, the estimated mean response operator $\hat{\mu}_{g_n}(x) \equiv \hat{\sfA}_{g_n} + \hat{\BB}_{g_n}x$, $x \in \HH$, is from \eqref{eqMRest} with another truncation $g_n$ and serves as the true mean response operator in the bootstrap world.
It is important to note that both original $\{ Y_i \}_{i=1}^n$ and bootstrap $\{ Y_i^*\}_{i=1}^n$ responses are conditionally generated given the same regressors $\{X_i\}_{i=1}^n$. 
Finally, the bootstrap estimators $\hat{\sfA}_{h_n}^*, \hat{\BB}_{h_n}^*$ are calculated in a similar manner to \eqref{eqFPCRest} as 
\begin{align} \label{eqFPCRestBTS}
	\hat{\BB}_{h_n}^* \equiv \hat{\Dt}_n^* \hat{\ga}_{h_n}^{-1}, \quad \hat{\sfA}_{h_n}^* \equiv \bar{Y}^* - \hat{\BB}_{h_n}^* \bar{X}
\end{align}
with the original truncation $h_n$,
where 
\[
\hat{\Dt}_n^* \equiv n^{-1} \sum_{i=1}^n \{(Y_i^* - \bar{Y}^*) \otimes (X_i - \bar{X})\}
\]
is the cross-covariance operator between regressors and bootstrap responses with $\bar{X} \equiv n^{-1} \sum_{i=1}^n X_i$ and $\bar{Y}^* \equiv n^{-1} \sum_{i=1}^n Y_i^*$. 
The bootstrap mean response operator $\hat{\mu}_{h_n}^*$ is then 
\begin{align} \label{eqMRestBTS}
	\hat{\mu}_{h_n}^*(x) \equiv \hat{\sfA}_{h_n}^* + \hat{\BB}_{h_n}^* x, \quad x \in \HH.
\end{align}
The last truncation $h_n$ involves both original and bootstrap estimators.
Such multiple tuning parameters are often considered in previous resampling works \cite[cf.][]{FH92, YDN23RB}.

Regarding the initial truncation $k_n$, a consistency of the corresponding regression estimators are sufficient to justify the above residual bootstrap; this guarantees consistent bootstrap error distribution by Theorem~S4 
of the online supplementary material.
For the sake of completeness, we provide a formal consistency result of the regression estimators in Theorem~S1 
of the online supplementary material.

The other extra truncation $g_n$ plays a more significant role in bootstrap inference, requiring additional conditions.
The truncation $g_n$ not only necessitate conditions for consistent estimators (cf.~Theorem~S1 
of the online supplementary material)
but we also need the following relative growth condition:
\begin{enumerate}[(R)]
	\item $\tau \equiv \lim_{n\to\infty} h_n/g_n \geq 1$. \label{condRatio}
\end{enumerate}
Condition~\ref{condRatio} illustrates the asymptotic relationship between the main truncation levels $h_n,g_n$; the second truncation $g_n$ cannot be asymptotically larger than the first truncation $h_n$. 
If this condition does not hold (i.e., $\tau \in (0,1)$), even when the random function $X$ is Gaussian, the residual bootstrap may be invalid.
Heuristically speaking, this is because the scaling with the first truncation $h_n$ is not sufficiently large to neglect the bootstrap bias term associated to the second truncation $g_n$. 
See Theorem~S5 
of the online supplementary material for a specific case in which such failure can occur.

Using the residual bootstrap procedure described above, we approximate the conditional distribution of $T_n$ from \eqref{eqStat} given the regressors $\xx_n \equiv \{X_i\}_{i=1}^n$ and $X_0$ with the distribution of its bootstrap counterpart
\begin{align} \label{eqRBstat_bts}
	T_n^* = T_n^*(X_0) \equiv \sqrt{n \over \hat{t}_{h_n}(X_0)} \{\hat{\mu}_{h_n}^*(X_0) - \hat{\mu}_{g_n}(X_0)\}.
\end{align}
The bootstrap approximation is theoretically justified in the following theorem; the proof is deferred to Section~S4 
of the online supplementary material.

\begin{thm} \label{thmRB}
	Suppose that the assumptions of \autoref{thmCLT} hold along with Condition~\ref{condRatio}. 
	We also assume that (i) the truncation levels $k_n$ and $g_n$ satisfy Condition~\ref{condBasicTrunc} and (ii) Condition~\ref{condScale} holds for $g_n$. 
	Then, as $n\to\infty$, the residual bootstrap is valid for mean response $\hat{\mu}_{h_n}(X_0)$ in \eqref{eqMRest} in the sense that 
	\begin{align*}
		\pi(\hat{P}_n, P_n) \xrightarrow{\pr} 0,
	\end{align*}
	where
	$\hat{P}_n \equiv \pr(T_n^* \in \cdot|\yy_n, \xx_n, X_0)$ and $P_n \equiv \pr(T_n \in \cdot|\xx_n, X_0)$ denote the conditional distributions of $T_n^*$ and $T_n$, respectively. 	
\end{thm}

\section{Numerical studies} \label{sec5}

In this section, we evaluate the performance of the CLT in \autoref{sec3} and bootstrap in \autoref{sec4}.
Specifically, we approximate the coverage probabilities of confidence sets constructed by either CLT or bootstrap, 
which are detailed in \autoref{ssec_5_1}.
Upon describing simulation designs in \autoref{ssec_5_2},
we provide the empirical coverage probabilities in \autoref{ssec_5_3}.

\subsection{Asymptotic and bootstrap confidence sets} \label{ssec_5_1}

Both CLT and bootstrap consistency in Theorems~\ref{thmCLT}--\ref{thmRB} can provide important inferential tools in the FoFR model \eqref{eqFOF}. 
In particular, these can calibrate confidence sets for quantities related to mean response \citep{YDN23RB, YDN24PB, CMS07, CM13}.
We elaborate constructing such confidence sets in this section.
Since it is well-studied by \cite{YDN23RB, YDN24PB} that CLT intervals may significantly underestimate a given coverage rate $1-\ap$ in the SoFR \eqref{eqSOF}, the inference in the FoFR \eqref{eqFOF} could be also benefited from bootstrap.
Before numerical evaluation of the confidence sets, we provided the detailed constructions. 

We first consider confidence set for the mean response $\mu(X_0)$ itself. 
By continuous mapping theorem \cite[Theorem~2.7]{bill99}, the limiting distribution of the squared norm $\|T_n\|^2$ of $T_n$ in \eqref{eqStat} is given as the weighted sum of the chi-square variables $\sum_{l=1}^\infty \s_l V_l$, where $\s_l$ denotes the $l$-th eigenvalue of $\sa$ and $\{V_l\}_{l=1}^\infty$ is a sequence of independent chi-square random variables with 1 degree of freedom.
This can provide a way of constructing a confidence set for the mean response $\mu(X_0)$. 
Since $\sa$ and its eigenvalues $\{\s_l\}_{l=1}^\infty$ are still unknown, 
one could estimate them using an average of squared residuals (cf.~Theorem~S2 
of the online supplementary material):
\begin{align} \label{eqSaHat}
	\hat{\sa}_{k_n} \equiv n^{-1} \sum_{i=1}^{k_n}[\{Y_i - \hat{\mu}_{k_n}(X_i)\}  \otimes \{Y_i - \hat{\mu}_{k_n}(X_i)\}].
\end{align}
With sufficiently large integer $L$ for approximating the infinite sum  $\sum_{l=1}^\infty \s_l V_l$, the critical value can be obtained from the distribution of $\hat{V} \equiv \sum_{l=1}^L \hat{\s}_l V_l$, where $\hat{\s}_l$ denotes the $l$-th eigenvalue of the estimated error covariance $\hat{\sa}_{k_n}$.
We take $L = 10,000$ for our numerical studies. 
Then, the critical value $\hat{c}_{\mathrm{CLT},1-\ap}$ for the CLT confidence region is defined as the $1-\ap$ quantile of $\hat{V}$ as $\hat{c}_{\mathrm{CLT},1-\ap}$.

However, a bootstrap counterpart does not require such an additional estimation for $\sa$. 
To explain the bootstrap confidence region for the mean response $\mu(X_0)$, let $\hat{c}_{\mathrm{RB}, 1-\ap}$ denote the $1-\ap$ quantile of the distribution of the bootstrap squared norm $\|T_n^*\|^2$ of $T_n^*$ in \eqref{eqRBstat_bts}.
The bootstrap critical value $\hat{c}_{\mathrm{RB}, 1-\ap}$ can be approximated by a Monte Carlo method. 
With either critical values, we can build confidence \emph{balls} for the mean response $\mu(X_0)$ as 
\begin{align} \label{eqCR}
	\widehat{CR}_{m, \mathrm{MR}} \equiv \left\{ y \in \HH:  \|\hat{\mu}_{h_n}(X_0) - y\| \leq  \hat{c}_{m,1-\ap} \sqrt{\hat{t}_{h_n}(X_0) \over n}  \right\}
\end{align}
for $m \in \{\mathrm{CLT}, \mathrm{RB}\}$.

Next, we construct confidence sets for lower dimensional quantities stemming from the mean response: projection and evaluation \cite[cf.][Corollarires~10--11]{CM13}.
First, for a general Hilbert space $\HH$, continuous mapping theorem \cite[Theorem~2.7]{bill99} suggests that the projection $\langle T_n(X_0), x \rangle$ converges in distribution to the centered normal distribution with variance $\langle \sa x, x \rangle$. This provides a confidence interval for the projection $\langle \mu(X_0), x \rangle$.
For evaluation, we may consider a more specific Hilbert space $\HH = \W_0^{2,1}([0,1]) \equiv \{ f \in L^2([0,1]): f(0)=0, f' \in L^2([0,1]) \}$ with inner product $\langle f, g, \rangle = \int_0^1 f'(t) g'(t) dt$ for $f,g, \in \W_0^{2,1}([0,1])$. 
In this case, since the evaluation functional $\W_0^{2,1}([0,1]) \ni f \mapsto f(t) \in \R$ for each $t \in [0,1]$ is continuous, the evaluation $\{T(X_0)\}(t)$ of the statistic  is asymptotically normal with limiting variance $\sa(t,t)$ \cite[cf.][Theorem~2.7 again]{bill99}, leading a confidence interval for the evaluation $\{\mu(X_0)\}(t)$ of the mean response. 
Upon replacing $\sa$ with its sample counterpart $\hat{\sa}_{k_n}$ given in \eqref{eqSaHat},
the critical values of the confidence intervals for the projection $\langle \mu(X_0), x \rangle$ and evaluation $\{\mu(X_0)\}(t)$ are given as 
\begin{align*}
	\hat{c}_{\mathrm{CLT},\mathrm{proj},x,1-\ap} 
	& =  \langle \hat{\sa}_{k_n} x, x \rangle z_{1-\ap/2},\\
	\hat{c}_{\mathrm{CLT},\mathrm{eval},t,1-\ap} 
	& = \hat{\sa}_{k_n}(t,t) z_{1-\ap/2},
\end{align*}
where $z_{1-\ap/2}$ is the $1-\ap/2$ quantile of the standard normal distribution.
The bootstrap critical values $\hat{c}_{\mathrm{RB},\mathrm{proj},x,1-\ap}$ and $\hat{c}_{\mathrm{RB},\mathrm{eval},t,1-\ap}$ are then defined, without estimating $\sa$, respectively by the $1-\ap$ quantiles of the distributions of the absolute bootstrap projection and evaluation statistics
\begin{align*}
	|\langle T_n^*(X_0),x \rangle|
	& = \sqrt{n \over \hat{t}_{h_n}(X_0)} |\langle \hat{\mu}_{h_n}^*(X_0), x \rangle - \langle \hat{\mu}_{g_n}(X_0), x \rangle |, \\
	|\{T_n^*(X_0)\}(t)|
	& = \sqrt{n \over \hat{t}_{h_n}(X_0)}|\{\hat{\mu}_{h_n}^*(X_0)\}(t) - \{\hat{\mu}_{g_n}(X_0)\}(t)|.
\end{align*}
Finally, the asymptotic and bootstrap confidence intervals for the projection $\langle \mu(X_0), x \rangle$ and evaluation $\{\mu(X_0)\}(t)$ can be constructed as
\begin{align}
	\widehat{CI}_{m, \mathrm{proj}, x} 
	& \equiv \langle \hat{\mu}_{h_n}(X_0), x \rangle \pm \hat{c}_{m,\mathrm{proj},x,1-\ap} \sqrt{\hat{t}_{h_n}(X_0) \over n}  , \label{eqCIproj} \\
	\widehat{CI}_{m, \mathrm{eval}, t} 
	& \equiv \{\hat{\mu}_{h_n}(X_0)\}(t) \pm \hat{c}_{m,\mathrm{eval},t,1-\ap} \sqrt{\hat{t}_{h_n}(X_0) \over n} , \label{eqCIeval}
\end{align}
for $m \in \{\mathrm{CLT}, \mathrm{RB}\}$.

%

\subsection{Simulation setups} \label{ssec_5_2}

The regressor $\{X_i\}_{i=1}^n$ and error $\{\e_i\}_{i=1}^n$ functions are independently generated based on the following (truncated) Karhunen--Lo\'{e}ve expansion:
\begin{align*}
	Z = \sum_{j=1}^{J_0} \g_j^{1/2} \xi_j \phi_j,
\end{align*}
where $J_0 = 20$. 
To consider random functions with non-normal and dependent FPC scores, 
we set $\xi_j = \xi W_j$, 
where $\{W_j\}_{j=1}^\infty$ is a sequence of independent standard normal random variables,
which is independent of the latent random variable $\xi$. 
The latent variable $\xi$ is distributed as either the standard normal distribution $\nd(0,1)$ or the centered exponential distribution $\mathsf{Exp}(1)-1$ with rate 1. 
The eigenvalues are obtained based on the eigengaps $\dt_j \equiv \g_j - \g_{j+1} = 2j^{-a}$ with $\g_1 \equiv  2 \sum_{j=1}^\infty j^{-a}$, where $a \in \{2.5, 5\}$ represents the eigen decay rate. 
We vary $\xi$ and $a$ to see their effects.

We consider two distinct eigenfunctions for regressors and errors, 
which is adapted from \cite{YK25_2sofr}, 
given as follows.
For the orthonormal system to generate the regressor functions, we consider the set $\{t^j\}_{j=1}^{J_0}$ of the monomials, 
and orthonormalize it to obtain $\{\phi_{\mathrm{mono},j}\}_{j=1}^{J_0}$.
The error functions lie on the orthonormal system constructed from the (shifted) Chebyshev polynomials. 
For this, we consider the Chebyshev polynomials
$\{f_{\mathrm{Cheb},j}\}_{j=1}^{J_{\mathrm{true}}}$
of the first kind determined by the recurrence
formula $f_{\mathrm{Cheb},j}(s) =
2sf_{\mathrm{Cheb},j-1}(s) + f_{\mathrm{Cheb},j-2}(s)$ for $j \geq 2$
with $f_{\mathrm{Cheb},1}(s)=s$ and $f_{\mathrm{Cheb},0}(s)=1$
where $s \in [-1,1]$ \cite[Chapter~3]{Caro98}; these can be computed using \texttt{chebPoly} function in \texttt{pracma} R package.
We shift them to $[0, 1]$ so as to have $f_{\mathrm{Cheb}, \mathrm{shift},j}(t) = f_{\mathrm{Cheb},j}(2t-1)$ and use their orthonormalized version $\{\phi_{\mathrm{Cheb},j}\}_{j=1}^{J_0}$.
For orthonormalization of both systems,
we apply the Gram--Schmidt orthogonalization in \citet[Theorem~2.4.10]{HE15}.
These choices of orthonormal systems are fixed before running the simulation. 

For the slope operators, we consider the following three Hilbert--Schmidt operators, 
which are motivated by the settings in \cite{DT24}. 
To describe the first two slope operators under consideration, we define the sequences
\begin{align} \label{eqSlopeCoef}
	\beta_{1j} \equiv 3W_{\beta,1,j}j^{-b_1}, \quad
	\beta_{2j} \equiv W_{\beta,2,j}j^{-b_2},
\end{align}
with $b_1 = 1.5$ and $b_2 = 2.5$.
Here, both $W_{\beta,1,j}$ and $W_{\beta,2,j}$ are iid random variables that have 1/2 probabilities on either -1 or 1 so as to determine the signs of $\beta_{1j}, \beta_{2j}$ (respectively).
Using the coefficients in \eqref{eqSlopeCoef} and the orthonormal set of the following $J_0 = 20$ trigonometric functions given as
\begin{align*}
	\phi_{\mathrm{tri},2m-1}(t) & = \sqrt{2}\sin(2m\pi t), \\
	\phi_{\mathrm{tri},2m}(t) & = \sqrt{2}\cos(2m\pi t),
\end{align*}
for $m=1,2,\dots,J_0/2$,
we define 
\begin{align*}
	\BB_{\mathrm{prod}} & \equiv \left( \sum_{j=1}^{J_0} \beta_{1j} \phi_j \right) \otimes \left( \sum_{j=1}^{J_0} \beta_{2j} \phi_j \right), \\
	\BB_{\mathrm{diag}} & \equiv \sum_{j=1}^{J_0} 2W_{\beta,j} j^{-(b_1+b_2)/2} (\phi_j \otimes \phi_j).
\end{align*}
The last slope operator $\BB_{\mathrm{exp}}$ is defined by the integral operator of a product of two decreasing exponential functions as follows: for $f \in L^2([0,1])$, 
\begin{align*}
	\{\BB_{\mathrm{exp}}(f)\}(t) \equiv \int e^{-(t+s)} f(s) ds, \quad t \in [0,1].
\end{align*}

Finally, the response functions $\{Y_i\}_{i=1}^n$ are generated by the FoFR model \eqref{eqFOFdata} with the zero intercept $\sfA=0$ and small sample sizes $n \in \{30, 50, 100\}$.
Our focus is more on small sample size performance of the CLT and bootstrap confidence sets,
because correct coverage probabilities in large sample size cases are anticipated from the previous studies such as \cite{YDN23RB}.
Both bootstrap resample size and Monte Carlo size are set to be 1,000.
For projection and evaluation,  
the cubic function $x(t) = 10t^3 - 15t^4 + 6t^5$ and $t = 0.9$ are considered, respectively.

Regarding the truncation levels $k_n, h_n, g_n$, we adapt the recommendation of \cite{YDN23RB}.
We first choose the initial truncation $k_n$ using the leave-one-out cross validation.
As mentioned in this previous study, we set $g_n = k_n$ as there was no big benefit from $g_n$ different than $k_n$.
Also, since it was already observed that slightly larger $h_n$ than $g_n$ provides better coverage accuracy, we only consider a few candidates $h_n = k_n + \dt$ with $\dt \in \{0,1,2\}$.

\subsection{Empirical coverage rates} \label{ssec_5_3}

In the simulation designs given in \autoref{ssec_5_2}, we assess the confidence sets in \eqref{eqCR}--\eqref{eqCIeval} by computing the empirical coverage probabilities,
specifically the proportions in which confidence sets contain the true objects: mean response $\mu(X_0)$, its projection onto $x$, or its evaluation $\{\mu(X_0)\}(t)$. 
We then compare the coverage rates of the confidence sets \eqref{eqCR}--\eqref{eqCIeval} from CLT and residual bootstrap in the FoFR model \eqref{eqFOF}.
Since the results are similar across all scenarios,
we present only partial results in \autoref{tb_cover},
using $a = 2.5$ and $\xi \sim \mathsf{Exp}(1)-1$ for generating both regressor and error functions and $\BB = \BB_{\mathrm{prod}}$.

\begin{table*}[!b]
	\centering
	\caption{
		\small Empirical coverage rates of both CLT and bootstrap (abbreviated as RB) confidence sets when $\ap=0.05$.
		The coverages within [0.930, 0.970] are in bold.
		The abbreviations MR, proj, and eval stand for mean response $\mu(X_0)$,  projection $\langle \mu(X_0), x \rangle$, and evaluation $\{\mu(X_0)\}(t)$, respectively.
	}
	\label{tb_cover}
	\renewcommand{\arraystretch}{1.2}  
	
	\begin{tabular*}{0.6\textwidth}{c|c|ccc|ccc}
		&    Methods    & \multicolumn{3}{c|}{CLT} & \multicolumn{3}{c}{RB}  \\ \hline
		$g_n = k_n$ &    $n$    & $30$    & $50$    & $100$    & $30$    & $50$    & $100$    \\ \hline
		\multirow{3}{*}{$h_n=k_n$} & MR & 0.503 & 0.587 & 0.648 & 0.780  & 0.853 & 0.895 \\
		& eval   & 0.719 & 0.750  & 0.795 & 0.894 & 0.928 & \textbf{0.943} \\
		& proj   & 0.639 & 0.729 & 0.766 & 0.805 & 0.875 & 0.911 \\ \hline
		\multirow{3}{*}{$h_n=k_n+1$} & MR & 0.629 & 0.682 & 0.705 & 0.847 & 0.885 & 0.920  \\
		& eval   & 0.779 & 0.807 & 0.819 & 0.897 & \textbf{0.933} & \textbf{0.959} \\
		& proj   & 0.726 & 0.782 & 0.793 & 0.849 & 0.893 & \textbf{0.930}  \\ \hline
		\multirow{3}{*}{$h_n=k_n+2$} & MR & 0.702 & 0.736 & 0.754 & 0.881 & 0.906 & \textbf{0.939} \\
		& eval   & 0.805 & 0.827 & 0.836 & 0.905 & \textbf{0.937} & \textbf{0.952} \\
		& proj   & 0.780  & 0.817 & 0.832 & 0.883 & 0.910  & \textbf{0.943}
	\end{tabular*}
\end{table*}

The key finding is that the residual bootstrap performs well in general and even with a sample size of $n=30$, 
achieving up to 90\% of coverage with its confidence sets.
In contrast, CLT confidence sets tend to undercover the true objects,
with coverage rates not exceeding 85\% even for $n=100$. 
Among three types of confidence sets, the evaluation interval \eqref{eqCIeval} provides the highest coverages,
whereas the confidence ball \eqref{eqCR} for the whole mean response $\mu(X_0)$ appears to struggle to reach the nominal level $\ap = 0.05$. 
All confidence sets demonstrate the asymptotic results in Theorems~\ref{thmCLT}--\ref{thmRB} because coverage rates become closer as the sample sizes increases.
Finally, as shown by \cite{YDN23RB} for SoFR, bootstrap confidence sets for FoFR with larger $h_n$ than $g_n$ generally yield higher coverages.

%
%
%
%
%
%

%
%
%
%
%
%
%
%
%
%

\section{Real data example} \label{secData}

We illustrate bootstrap inference for FoFR, 
supported by the theoretical validity in \autoref{thmRB} and the numerical results in \autoref{sec5}. 
The focus of this section is on the Canadian weather dataset \citep{RS05}, available in the R package \texttt{fda} \citep{fda},
which has been frequently studied by \cite{Lian15, SDWM18, FGS16, LJLZ19, CFLV22}  in FoFR contexts.
The main goal of our analysis is to conduct inference on the mean responses,
a direction that has not been considered in these previous works.

\begin{figure*}[b!]
	\centering
	\includegraphics[width=0.79\linewidth]{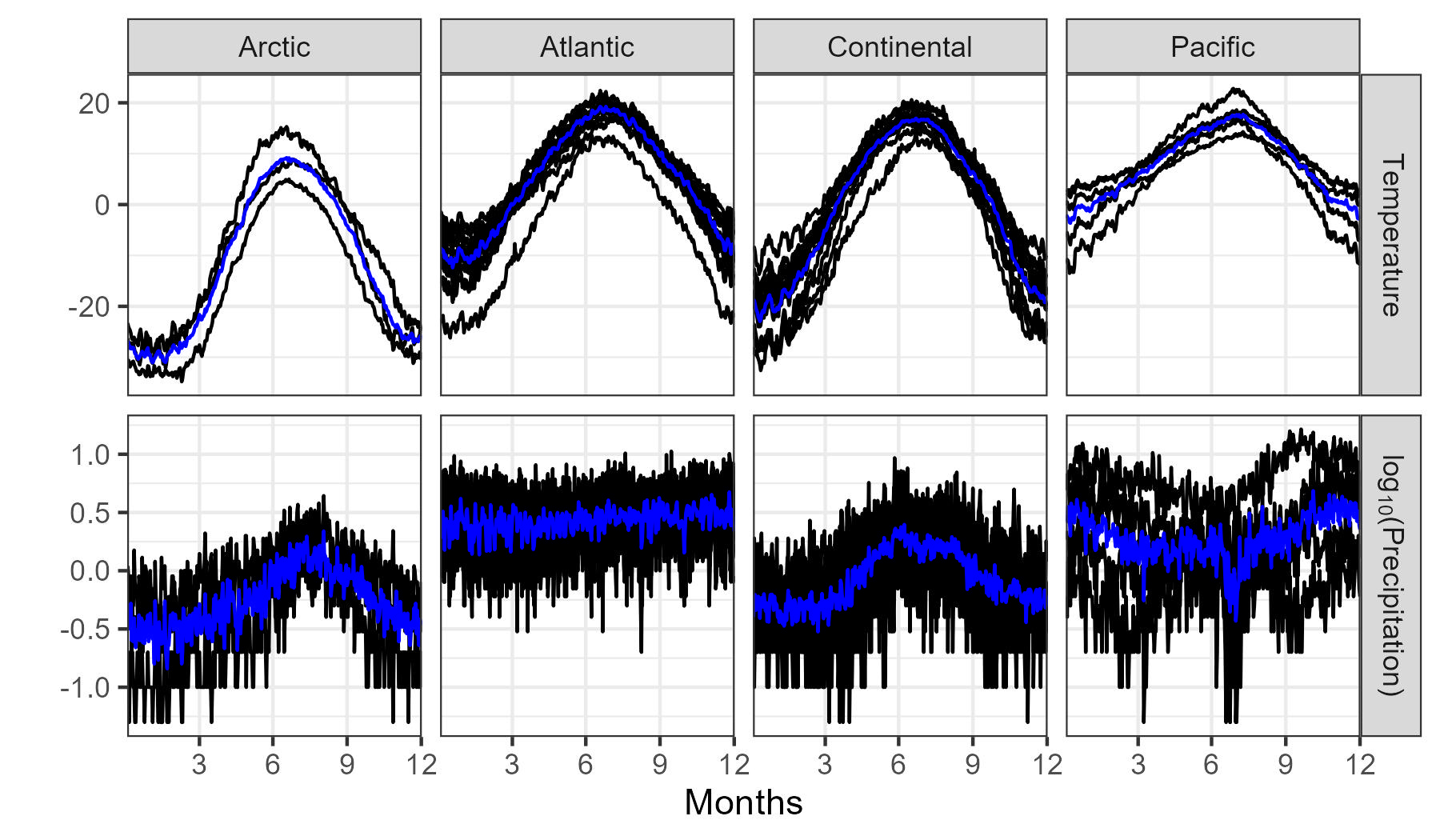}
	\caption{
		Canadian weather dataset.
		Averaged daily temperatures (upper) and (log) precipitations (lower) over a year  for four regions in Canada,
		where the regional average curves are colored in blue.
	}
	\label{figRDAdata}
\end{figure*}

The temperatures (in $^{\circ}$C) and precipitations (in mm) were recorded daily at 35 Canadian cities from 1960 to 1994,
and the dataset \texttt{CanadianWeather} stores the average values  across years, at each day and for each city. 
In our analysis, we use daily log-transformed precipitation (base 10) as the responses $\{Y_i\}_{i=1}^n$ and daily temperatures as the regressors $\{X_i\}_{i=1}^n$. 
The dataset categorizes cities into four distinct regions—Arctic, Atlantic, Continental, and Pacific,
which we use to construct new regressors by averaging temperature curves across cities within each region.
The functional observations are displayed in \autoref{figRDAdata} along with these regional averages highlighted in blue. 
We select $k_n=5$ via leave-one-out cross validation, and set $g_n=k_n=5$ and $h_n=k_n+1 = 6$, as these choices perform well in our numerical studies. 
For bootstrap, we generate 1,000 resamples. 

Since the bootstrap confidence ball $\widehat{CR}_{\mathrm{RB},\mathrm{MR}}$ in \eqref{eqCR} is difficult to visualize directly,
we reformulate the problem as a hypothesis test for the null hypothesis $H_0:\mu(X_0) = \eo[Y]$,
which posits that the mean response $\mu(X_0)$ at a given region equals the global mean $\eo[Y]$.
The decision rule is to reject $H_0$
(i) if the global average response $\bar{Y}  \equiv n^{-1}\sum_{i=1}^n Y_i$ does not lie within $\widehat{CR}_{\mathrm{RB},\mathrm{MR}}$,
i.e., $\bar{Y}  \equiv n^{-1}\sum_{i=1}^n Y_i\notin \widehat{CR}_{\mathrm{RB},\mathrm{MR}}$,
or equivalently, (ii) if the following bootstrap p-value $\hat{p}_n$ is less than the significance level $\ap$:
\begin{align*}
	\hat{p}_n
	\equiv \pr^* \left( \|T_n^*(X_0)\|^2> n \|\hat{\mu}_{h_n}(X_0) - \bar{Y}\|^2 / \hat{t}_{h_n}(X_0) \right).
\end{align*}
This decision rule can be justified,
because under $H_0$, $\mu(X_0)$ in \eqref{eqStat} can be replaced by $\bar{Y}$ due to the following asymptotic relationship:
\begin{align*}
	& \sqrt{n \over \hat{t}_{h_n}(X_0)} \{\hat{\mu}_{h_n}(X_0) - \bar{Y}\}
	\\=& \sqrt{n \over \hat{t}_{h_n}(X_0)} \{\hat{\mu}_{h_n}(X_0) - \mu(X_0)\}
	+ \sqrt{n \over \hat{t}_{h_n}(X_0)} \{\eo[Y] - \bar{Y}\}
	\\=&  \sqrt{n \over \hat{t}_{h_n}(X_0)} \{\hat{\mu}_{h_n}(X_0) - \mu(X_0)\} + O_\pr(t_{h_n}(X_0)^{-1/2}).
\end{align*}
The resulting bootstrap p-values are 0 for Arctic, Atlantic, and Continental regions, and 0.008 for the Pacific region.
These results provide strong evidence against the null hypothesis in all cases, 
suggesting that each regional mean response $\mu(X_0)$ differs significantly from the global mean $\eo[Y]$.

To illustrate bootstrap inference for projection and evaluation,
we consider the constant function $x=1$ and evaluation time $t=359$.
Then, the projection $\langle \mu(X_0), x \rangle$ represents the overall average log precipitation in each region,
while $\{\mu(X_0)\}(t)$ stands for the log precipitation on the Christmas day. 
We compute the bootstrap confidence intervals in \eqref{eqCIproj}--\eqref{eqCIeval} and summarize them in \autoref{tbRDAintervals}. 
We obtain similar ANOVA-type interpretations by comparing these intervals with the sample averages: $\langle \bar{Y}, x \rangle = 0.173$ or $\bar{Y}(t) = 0.238$.
If either of these values falls outside the corresponding bootstrap intervals in \autoref{tbRDAintervals},
we conclude that
the regional mean log precipitation --- 
either the overall average $\langle \mu(X_0), x \rangle$ or the Christmas day evaluation $\{\mu(X_0)\}(t)$
--- is not equal to its global counterpart $\langle \eo[Y],x \rangle$ or $\eo[Y](t)$, respectively.
This happens for all regions and for both projections and evaluation,
except for the case with projection in the Pacific region. 

\begin{table}[t!]
	\centering
	\caption{
		Bootstrap intervals for projection $\langle \mu(X_0), x \rangle$ and evaluation $\{\mu(X_0)\}(t)$ of the mean response $\mu(X_0)$ from the Canadian weather dataset,
		where $x$, $t$, and $X_0$ are given as in the text. 
		The target quantities that the intervals would include are $\langle \bar{Y}, x \rangle = 0.173$ and $\bar{Y}(t) = 0.238$ for projection and evaluation, respectively.
		The intervals containing the corresponding target quantity are indicated in italics.
	}
	\label{tbRDAintervals}
	
	\renewcommand{\arraystretch}{1.2}  
	
	\begin{tabular}{c|cc}
		& Projection & Evaluation \\ \hline
		Arctic & [-0.395,	-0.093] &	[-0.745,	-0.263] \\
		Atlantic & [~0.354,	~0.498]	&   [~0.558,	~0.790] \\
		Continental & [-0.150,	~0.003]	&   [-0.353,	-0.093] \\
		Pacific & \textit{[~0.152,	~0.366]}	&   [~0.311,	~0.654]
	\end{tabular}
\end{table}

\section{Conclusions and outlook} \label{sec6}

In this work, we refine the existing CLT and develop a residual bootstrap method for mean response in FoFR models. 
The generalized CLT is indeed applicable, conditionally on all regressors, to the cases with more varied regressors  (e.g., those with non-normal and dependent FPC scores) or the cases where a slope operator do not require absolute summability.
Also, upon providing its consistency with rigor, we found that the proposed residual bootstrap can lead to higher numerical accuracy of the subsequent confidence sets for mean response or associated scalar quantities.
A few direct extensions are listed below, which we do not discuss further in this paper.

\begin{enumerate}[(i)]
	
	\item 
	The new regressor $X_0$ is not necessarily restricted to the conditions imposed in this paper. 
	For example, $X_0$ can be dependent on and non-identical to data regressors \citep{YDN23RB} or it can be a fixed element \citep{CMS07, KH16a}, where the latter case may need a different technique. 
	
	\item 
	Analogously to \cite{YDN23RB}, one could consider predicting the new response $Y_0$ at the new regressor $X_0$.
	In this context, bootstrap prediction is expected to provide better accuracy than CLT counterpart for small sample sizes.
	
	\item 
	Simultaneous inference (or prediction) for mean responses at multiple new regressors $\{X_{0l}\}_{l=1}^L$ is another benefit of bootstrap,
	as it is intractable with CLTs due to the potential dependence between mean response estimators \citep{YDN23RB, YDN24PB}.
	Bootstrap for FoFR allows us to conduct inference for more complex situation, such as finding simultaneous intervals for multiple mean response functions $\{\mu(X_{01})\}(t_1), \dots, \{\mu(X_{0L})\}(t_L)$ evaluated at possibly distinct time points $t_1, \dots, t_L$. 
\end{enumerate}

Our asymptotic results will have significant impacts, paving the way multiple (non-trivial) avenues for future research.
We conclude this paper with several open questions and potential directions.

\begin{enumerate}[(a)]
	
	\item 
	Heteroscedastic errors were considered for SoFR  \citep{DHA09, YDN24PB}, but to our knowledge, there has been no study of heteroscedastic FoFR.
	In this more complex situation, a new CLT is necessary, possibly a novel scaling term, and a different bootstrap approach would be more appropriate. 
	
	\item 
	Another main FoFR estimation approach is a method based on reproducing kernel Hilbert spaces (RKHS) framework \cite[e.g.,][]{SDWM18, DT24}.
	This RKHS-based estimation may have benefits from its continuous regularization (among other merits) compared to the FPCR estimation;
	notably, the resulting slope estimator has its own central limit theorem. 
	A drawback of the RKHS-based estimation could be that it necessarily restricts the parameter space to a smaller subspace in $\HH$.
	Another representative approach for functional linear regression is ridge estimation \citep{HH07, BCF17}.
	Despite the existence of various methods, there has been no comprehensive comparative study covering both theoretical and numerical aspects, which would be invaluable for practitioners.
	
	%
	%

	\item 
	Hypothesis testing of the slope operator for FoFR is indeed a critical issue \citep{KMSZ08}.
	Given that a residual bootstrap for SoFR testing has shown promising results by \cite{KH24},
	it is plausible to expect similar benefits from using bootstrap for FoFR testing.

	\item 
	
	Functional data may be observed incompletely, leaving numerous missing values,
	often referred to as sparse functional data \citep{YMW05b}. 
	This sparseness present unique challenges.
	In such cases, analogously to findings for the dense FoFR by \cite{CM13}, it would be valuable to establish CLTs for either a slope estimator or its associated mean response. 
	Additionally, a subsequent bootstrap method could be useful for calibrating confidence sets even when observing sparse functional data.
	This remains an open problem to date.
	
	\item
	In functional regression, many estimation approaches involve tuning parameters; 
	however, selecting optimal tuning parameters for inferential purposes has not been thoroughly studied. 
	The optimal choice of truncation levels in FPCR estimators may vary depending on the specific inferential objectives and the bootstrap scheme employed,
	as indicated by prior work in other resampling contexts \citep{Lahiri99, NL04, Nord09}. 
	A comprehensive investigation of this issue is beyond the scope of the present study and is left for future research.


	%
	%
	%
		%
		%
	
\end{enumerate}



\section*{Supplementary material}

The online supplementary material includes technical proofs for all theoretical results in the main paper.


\bibliographystyle{imsart-nameyear}
\bibliography{RBinFLRM_FOF_bibfile}          
%
%

\end{document}